\documentclass[%
 aip,
 jmp,%
 amsmath,amssymb,
 preprint,%
]{revtex4-1}

\usepackage{graphicx}
\usepackage{dcolumn}
\usepackage{bm}

\begin{document}

\preprint{AIP/123-QED}

\title{Ballistic and Diffuse Electron Transport in Nanocontacts of Magnetics }%

\author{R.G. Gatiyatov }%
 \email{Ruslan.Gatiyatov@gmail.com}
 \affiliation{Zavoisky Physical-Technical Institute RAS, 420029 Kazan, Russia}%
\author{V.N. Lisin}
\author{A.A. Bukharaev}
\date{\today}
\begin{abstract}
The transition from the ballistic electron transport to the diffuse one is experimentally observed in the study of the magnetic phase transition in Ni nanocontacts with different sizes. It is shown that the voltage $U_C$ needed for Joule heating of the near-contact region to the critical temperature does not depend on the contact size only in the diffuse mode. For the ballistic contact it increases with decrease in the nanocontact size. The reduction of the transport electron mean free path due to heating of NCs may result in change of the electron transport mode from ballistic to diffusive one.
\end{abstract}

\maketitle
In modern micro- and nanoelectronics devices, as a rule, the diffuse electron transport is implemented when the transport mean free path $l_{tr}$ is less than the characteristic size of a device $d$. Further, with decrease in $d$ the ballistic electron transport becomes a typical one ($d << l_{tr}$) when electrons pass through a nanoobject without collisions. In metal nanocontacts (NCs) the condition $d << l_{tr}$ can hold at room and higher temperatures~\cite{Gatiyatov2010}. It was recently concluded upon studying the passage of electrons through NCs of magnetic metals~\cite{Gatiyatov2010, Chen} that not only in the diffuse NCs but also in the ballistic ones it is necessary to take into account Joule heating of the near-contact region. Moreover, the temperature of the ballistic NCs can reach the critical value at which the magnetic phase transition occurs. We established earlier~\cite{Gatiyatov2010} that the external voltage value $U_C$ necessary for heating the ballistic NCs to the phase transition point $T_C$ increases with reduction of the NC diameter $d$ while in Ref.~\cite{Chen} it was concluded that $U_C$ does not depend on $d$. In our opinion, it is interesting to clarify the origin of this contradiction since the study of the features of heating of NCs and the neighboring regions by the current is one of the topical problems of studying magnetic NCs. In particular, this is connected with the implementation of the spin transfer torque effect in such structures~\cite{Ralph, Hatami}.

To clarify the origin of such a noticeable distinction between the conclusions made in Ref.~\cite{Gatiyatov2010, Chen}, in this work the electron transport in Ni NCs is studied in a wider range  of the NC sizes than in Ref.~\cite{Gatiyatov2010} (range of the resistances at zero voltage $R_0=13\div400$\,Ohm which in our estimates corresponds to the diameter values $d= 1.5\div 12$\,nm).

The measurements were performed at room temperature in a wide range of voltages applied to NCs. Ni NCs were formed in solution using an electrochemical method described in Ref.~\cite{Gatiyatov2010} in detail. The conductance was measured by the four-contact method with two digital multimeters, Agilent 34410a. Current-voltage (I-V) characteristics of NCs were recorded by applying a single triangular voltage pulse to the circuit. The pulse frequency varied from $10$ to $100$\,Hz. Ni NCs were formed in the nickel sulfate solution $0.25$\,Ì NiSO$_4$ + $0.5$\,Ì H$_3$BO$_3$ (working voltage $1 \div 1.4$\,ÂV) in the gap between two Ni microwires fixed on a nonconducting substrate. I-V curves were recorded in the bath with bidistilled water, which has the conductivity much smaller than the NC conductivity.

\begin{figure}[h!]
	\begin{minipage}[h]{0.5\linewidth}
		\center{\includegraphics[height=1\linewidth]{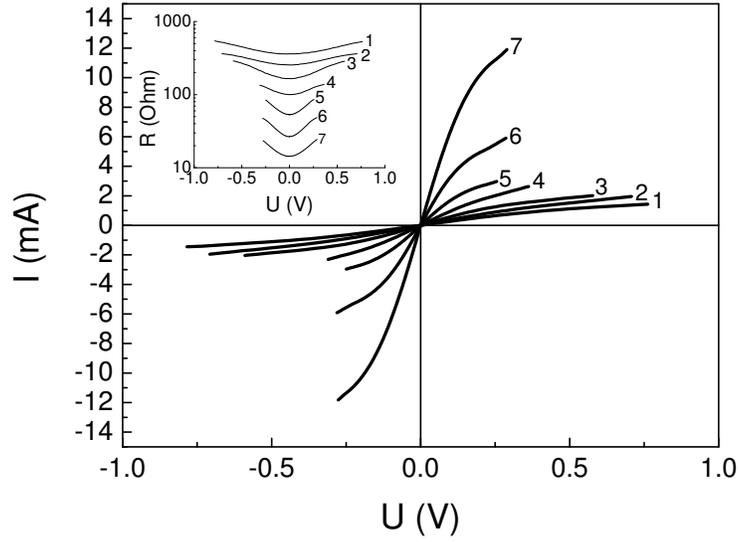}}\\(a)
	\end{minipage}
	\vfill
	\begin{minipage}[h]{0.5\linewidth}
		\center{\includegraphics[height=1\linewidth]{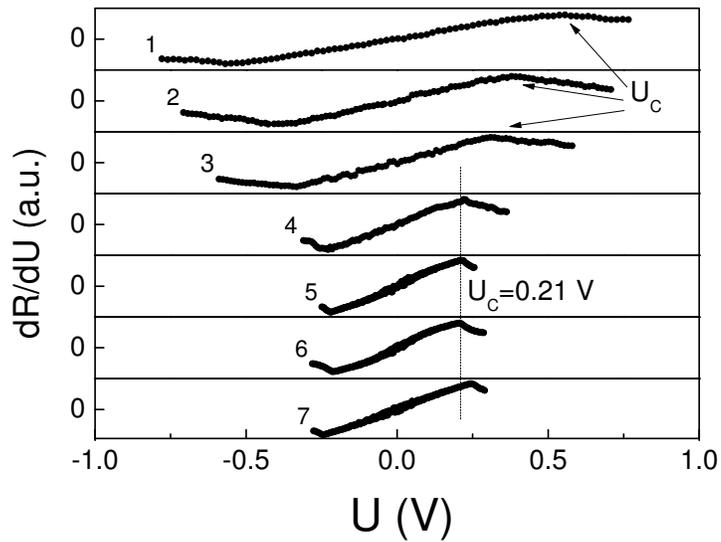}} \\(b)
	\end{minipage}	
\caption{I-V curves (a) and dependences of the resistance (inset) and voltage derivative of resistance on the applied potential difference (b) for Ni NC with different sizes. Numbers correspond to the following values of the NC resistance at zero potential difference: $1$ -- $363$\,Ohm, $2$ -- $254$\,Ohm, $3$ -- $165$\,Ohm, $4$ -- $100$\,Ohm, $5$ -- $53$\,Ohm, $6$ -- $26$\,Ohm, $7$ -- $14$\,Ohm.}
\label{fig1}
\end{figure}

Figure1 shows the I-V curves (Fig.1a) and dependences of the resistance $R=U/I$ (in the inset) and voltage derivative of resistance $dR/dU$ (Fig.1b) of Ni NCs with different diameters (different resistances $R_0$ at $U=0$\,V) on the value of the applied potential difference $U$. One can see in Fig.1a that the I-V curves are nonlinear. With increase in the applied voltage the NC resistance increases. The voltage derivative of resistance $dR/dU$ was found by the numerical differentiation of the experimental dependence $R(U)$. There are a maximum at $U=U_C$ and a symmetrically located minimum $U=-U_C$ in the dependences of the voltage derivative of resistance (Fig.1b). It is seen that with increase in the Ni NC size (reduction of the resistance $R_C=R(U_C)$) the $U_C$ value decreases and reaches a plateau for $R_C < 100$\,Ohm.

Figure 2 shows the experimental dependence of the $U_C$ value on $R_C^{-1}$. The $U_C$ value nonmonotonically declines with increase in $R_C^{-1}$ (increase in the contact size), and for $R_C < 100$\,Ohm reaches a plateau and becomes $U_C=0.21 \pm 0.03$\,V.

\begin{figure}[h!]
	\begin{minipage}[h]{0.5\linewidth}
		\center{\includegraphics[height=1\linewidth]{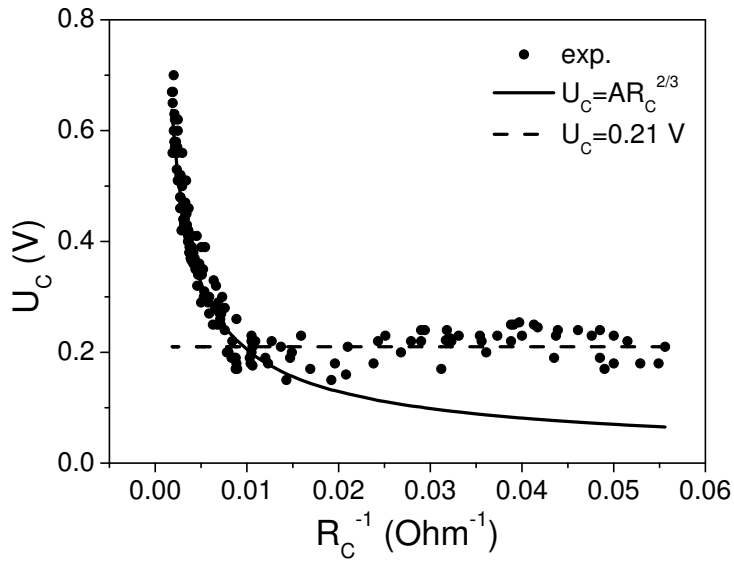}}\\(a)
	\end{minipage}	
\caption{Dependence of the voltage $U_C$ necessary for heating NC to the critical temperature on the NC resistance at this voltage (solid line, the dependence calculated according to equation~\eqref{eq:eq1}; dashed line corresponds to  $U_C = 0.21$\,V).}
\label{fig2}
\end{figure}

According to~\cite{Gatiyatov2010}, the nonlinearity of the I-V curves is due to heating of NC by a current running through it. It was shown that the maximum (minimum) in the $dR/dU(U)$ dependences (Fig.1b) is due to the local phase transition from the ferromagnetic to the paramagnetic state of the contact region because of heating above the Curie temperature~\cite{Gatiyatov2010}, which for Ni is $T_C = 631$\,K~\cite{Vonsovsky}.

We studied Ni NCs with $R_C > 100$\,Ohm in Ref.~\cite{Gatiyatov2010}. It was shown that in such NCs the ballistic electron transport was implemented. The dependence of the $U_C$ value on $R_C$ for this region is described by the expression:
\begin{eqnarray}\label{eq:eq1}
    U_C = A \cdot R_C^{\frac{2}{3}},
\end{eqnarray}
 where $A=(0.95\pm0.12)\cdot10^{-2}$\,V/Ohm$^\frac{2}{3}$. Figure 2 shows the dependence of $U_C$ on $R_C^{-1}$ (solid line according to~\eqref{eq:eq1}).

It should be emphasized that in the region $R_C < 100$\,Ohm the $U_C$ value no longer depends on the contact size within the scatter of the experimental values (Fig.2). It is known that in diffuse contacts the voltage value needed for heating to the same temperature does not depend on the contact size~\cite{Holm}, and is related to the temperature in the contact center by the following relation:
\begin{eqnarray}\label{eq:eq2}
    U_C^2 = 8   \int\limits_{T_0}^{T_C} \lambda \rho dT,
\end{eqnarray}
 where $\lambda$ is the coefficient of thermal conductivity, $\rho$ is the resistivity.

By numerically integrating~\eqref{eq:eq2}, using tabular data for the temperature dependence of the coefficient of thermal conductivity and resistivity of Ni~\cite{Handbook}, we find that $U_C=0.17$\,Ohm. The calculated value is somewhat below the average experimental value $U_C=0.21$\,V but is within the scatter of the experimental data. This allows us to state that for Ni NC with $R_C<100$\,Ohm the diffuse electron transport is implemented.

On the basis of the above one can conclude that the experimental data (Fig.2) indicate the transition from the ballistic ($d << l_{tr}(T_C)$) electron transport in Ni NC to the diffuse ($d > l_{tr}(T_C)$) one at the temperature of the magnetic phase transition $T_C$. In the ballistic mode $U_C$ depends on the NC size which contradicts the results of Ref.~\cite{Chen}.
       
In our opinion, this is due to the fact that the sizes of the studied Au/CoS$_2$ and Au/Ho contacts in Ref.~\cite{Chen} were too large for the implementation of the ballistic electron transport in NCs heated to the temperature of the phase transition. The estimates of $l_{tr}$  in Ref.~\cite{Chen} were given for the initial temperature $T_0=4.2$\,K. The increase in the resistance with increase in the voltage indicates the reduction of $l_{tr}$. This can lead to the transition from the ballistic electron transport ($l_{tr}(T_0) >> d$) to the diffuse ($l_{tr}(T_C) < d$) one. Using the tabular resistivity value $\rho(T_N)=52 \cdot 10^{-8}$\,Ohm$\cdot$m for Ho [Ref.~\cite{Vonsovsky}] at the Neel temperature $T_N$ and assuming  $\rho \cdot l_{tr} = 10^{-15}$\,Ohm$\cdot$m$^2$ the same as in Ref.~\cite{Chen}, we obtain  $l_{tr}(T_N)=1.9$\,nm. The estimated $l_{tr}(T_N)$ value is less than the contact sizes in Ref.~\cite{Chen} therefore in these contacts the diffuse electron transport is implemented. Analogous estimates can be made for the Au/CoS$_2$ contact.

Thus, one can consider that in Ref.~\cite{Chen} the electron transport in NCs heated to the temperature of the magnetic transition was rather diffuse than ballistic, since the $l_{tr}$ values were overestimated. This is also confirmed by the observation that the voltage value $U_C$  does not depend on the contact size which agrees well with the ideas about the diffuse electron transport in similar structures.

The growth in $U_C$  with decrease in the contact diameter can be qualitatively understood as follows. For ballistic NCs, the size of the region where heat is released does not depend on the contact size~\cite{Levinson}. The temperature of the heated region is determined by the released power $P = U^2/R$. Its resistance $R$ increases with decrease in the contact diameter. To keep the NC temperature constant, it is necessary for the released power to be constant as well and, consequently, it is necessary to increase the voltage $U$.

It is interesting to note that value of the transport electron mean free path at the temperature of the phase transition can be estimated from the experimental dependence of $U_C$ on $R^{-1}_C$. The NC resistance can be expressed by means of the Wexler interpolation formula~\cite{Wexler} via the Maxwell resistance $R_M$ for the diffuse contact and Sharvin resistance $R_{Sh}$, describing the NC resistance in the ballistic mode:
\begin{equation}\label{eq:eq3}
	R_C = \gamma_W R_M + R_{Sh} =\frac{\gamma_W \rho(T_C)}{d} + \frac{\gamma_{Sh} C}{d^2},
\end{equation} 	
where $\rho$  is the resistivity, $d$ is the NC diameter,  $C = \rho \cdot l_{tr}$ is the coefficient, which is determined by the properties of the Fermi surface and does not depend on the dissipation mechanism, $\gamma_W$ is the correction coefficient ~\cite{Wexler}, $\gamma_{Sh}=16 / 3 \pi$. By equalizing $\gamma_W R_M$ and $R_{Sh}$, we find that in the transition point:
\begin{equation}\label{eq:eq4}
d = \gamma_{Sh} l_{tr} / \gamma_W.
\end{equation}
The value of the contact resistance corresponding to the transition from the ballistic electron transport to the diffuse one is found from the intersection of the two curves (Fig.2), describing the ballistic and diffuse contacts, which is $R^*_C = 100$\,Ohm. By substituting $R^*_C$ and ~\eqref{eq:eq4} in~\eqref{eq:eq3}, we find $l_{tr}(T_C) = 1.8$\,nm, $d = 4.1$\,nm and $C = \rho \cdot l_{tr} = 5 \cdot 10^{-16}$\,Ohm$\cdot$m$^2$. The following parameter values: $\gamma_W = 0.75$ [Ref.~\cite{Wexler}],  $\rho(T_C) =27.8 \cdot 10^{-8}$\,Ohm$\cdot$m [Ref.~\cite{Handbook}] were used in calculations.

Thus, it was demonstrated that it is important to take into account the reduction of the transport electron mean free path due to heating of NCs upon studying the mode of the electron transport through NCs. This reduction of the electron mean free path can lead to the transition from the ballistic electron transport to the diffuse one. As a result, the voltage value needed for heating of NCs to the critical temperature no longer depends on the NC size.

This work is supported by the grant of the Russian Foundation for Basic Research and programs of the Department of Physical Sciences of the Russian Academy of Sciences.

We are grateful to S.A Ziganshina for the help in preparing working electrolytes and to I. A. Garifullin for useful remarks upon the discussion of results.
\end{document}